\documentclass[useAMS,usenatbib]{mn2e}
\usepackage{graphicx}	
\usepackage{lineno}	
	
\title[North-South Asymmetry in the Solar Coronal Rotation]{North-South Asymmetry in the Solar Coronal Rotation}

\author[Hari Om Vats and Satish Chandra]{Hari Om Vats$^{1}$\thanks{
vats@prl.res.in (HOV)} and Satish Chandra$^{2}$\thanks{satish0402@gmail.com (SC)}\\
$^{1}$Physical Research Laboratory, Ahmedabad - 380 009, INDIA.\\
$^{2}$Department of Physics, PPN College, Kanpur - 208 001, INDIA.}

\begin{document}

\date{Accepted for publication in MNRAS Letters}

\pagerange{\pageref{firstpage}--\pageref{lastpage}} \pubyear{8888}

\maketitle


\label{firstpage}


\begin{abstract}
The solar images at 17 GHz by Nobeyama Radio Heliograph and in X-ray by soft X-ray telescope (SXT) on board \textit{Yohkoh} satellite have been of particular interest for the estimation of solar coronal rotation using flux modulation approach. These studies established that the solar corona rotates differentially. The radio images estimate equatorial rotation period lower than those estimated by the X-ray images. The latitude profiles of the coronal rotation have temporal variability. It is very interesting that the space-time plots of sidereal rotation period revealed clearly North-South asymmetry. The asymmetry appears to change its sign in odd and even activity cycles of the Sun.
\end{abstract}


\section{Introduction}

Solar rotation has been known for the past four centuries. This was first seen by Gallilei using his first optical telescope. Studies on solar rotation both in the solar interior and its atmosphere attained great interest in the last four decades. The variation of characteristics of time-varying processes occurring in the solar atmosphere is tightly connected with prolonged large-scale manifestation of solar activity. The interaction between solar rotation, convection and magnetic fields indisputably is the reason for such activity. The study of the rotation of the solar interior is being carried out by Helioseismology [\citet{howe09} and references therein]. The rotation of solar photosphere, chromosphere and corona is investigated by ground based and satellite observations of the Sun. There are three basic methods used for these investigations, namely, tracers, spectroscopy and flux modulation \citep{howard96, beck00, vats01}. The initial studies of solar rotation observed the movement of sunspots as tracers across the visible disc. Thereafter many other quasi-permanent solar features were found to be useful as tracers for measuring solar rotations. These include plages, faculae, coronal holes, supergranules, giant cells, etc. [\citet{beck98} and references therein]. Measurements of the rotation rate by usage of these tracers yielded different numbers and profiles. Some of these measurements do match with each other whereas many do not match. Disk integrated radio flux at 2.8 GHz and at a few more frequencies have been used for studies on long term average variation in the solar corona \citep{vats98a, vats98b, kane01, chandra11}. These measurements revealed quite large variability of coronal rotation. Multi-frequency simultaneous measurements of solar radio flux indicated that the corona has a significant rotation gradient as a function of altitude \citep{vats01}.

The disk integrated measurements give no information on the latitude variations. Radio images at 17 GHz for the period 1999 - 2005 give very interesting information about the differential rotation of the solar corona at the height of these emissions. The radio images are from Nobeyama Radio Heliograph for the years 1999 - 2001 \citep{chandra09}. The X-ray images of \textit{Yohkoh} SXT for the years 1992 - 2001 are used by \citet{chandra10}. \citet{vats10} have discussed the temporal variation of rotation as a function of solar activity during the years 1999 - 2005 for radio observations and for the years 1992 - 2001 for X-ray measurements. This study showed that the equatorial part of the corona rotates slightly faster than that of the photosphere and chromosphere. The region of higher latitudes in the corona rotates much faster than the corresponding region of the lower solar atmosphere. \citet{casas06} confirmed that the solar surface rotates slowly at the equator during low solar activity period and it shows a higher level of differential rotation with latitude. A similar, but smaller decrease in the equatorial rotation rate between solar cycles 13 and 14 ($\sim$ 1901), accompanied by a change in the decay rate of sunspots was reported by \citet{balthasar86}. This epoch coincides with the minimum of the 80-year activity cycle. \citet{chandra10}, using X-ray images, suggested that the equatorial rotation rate is anti-correlated with the sunspot activity, except in the year 2000. \citet{chandra}, using radio images (for the years 1999 - 2005), at 17 GHz showed that the equatorial rotation rate and the sunspot number have a lag of 3 years. Furthermore, the differential part of the rotation is almost in phase with the sunspot number. Here, we present the space-time plots of the sidereal rotation period obtained from the observations of (1) \textit{Yohkoh} X-ray during 1992 - 2001 and (2) the radio images at 17 GHz during 1999 - 2005. The former partly covers solar cycles 22 and 23 whereas the latter is only during solar cycle 23.

\begin{figure}
\centering{
  \includegraphics[width=83mm]{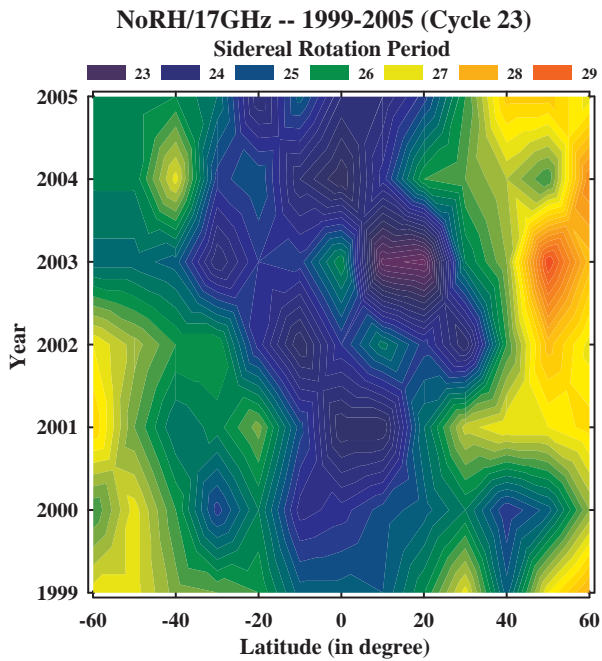}}
\caption{Space-time plots of sidereal rotation period from radio images at 17 GHz observed during 1999-2005. This covers partially the descending phase of solar cycle 23. The colour codes (shown on top) ranges between 23-29 days represent sidereal rotation period.}
\label{Figure1_NS}
\end{figure}

\begin{figure}
\centering{
  \includegraphics[width=83mm]{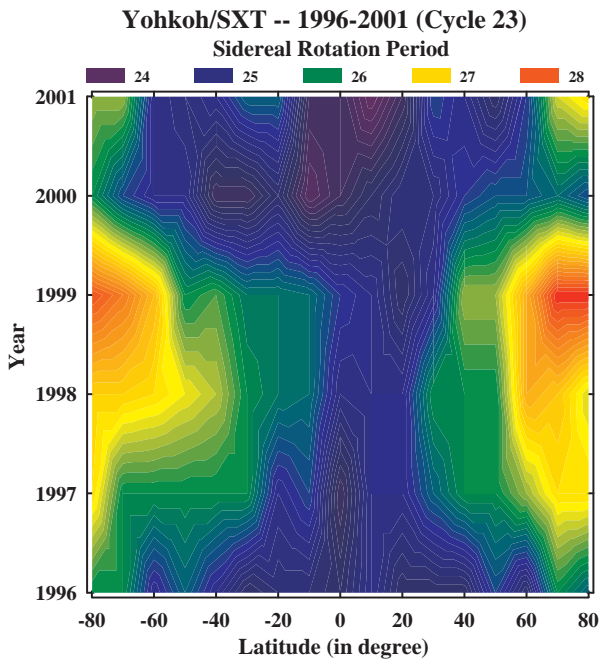}}
\caption{Same as Figure \ref{Figure1_NS} for X-ray images observed during 1996-2001. This covers partially the ascending phase of solar cycle 23. The colour codes (shown on top) ranges between 24-28 days represent sidereal rotation period.}
\label{Figure2_NS}
\end{figure}

\begin{figure}
\centering{
  \includegraphics[width=83mm]{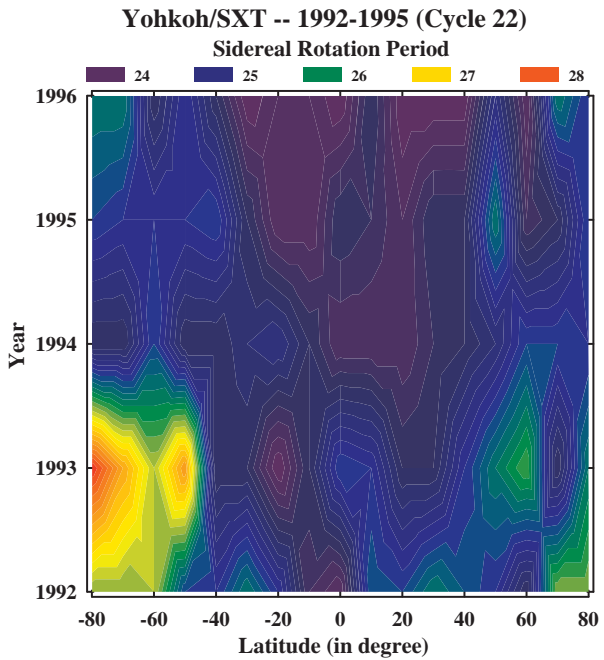}}
\caption{Same as Figure \ref{Figure2_NS} observed during 1992-1995, which covers partially the descending phase of solar cycle 22.}
\label{Figure3_NS}
\end{figure} 

\begin{figure}
\centering{
  \includegraphics[width=83mm]{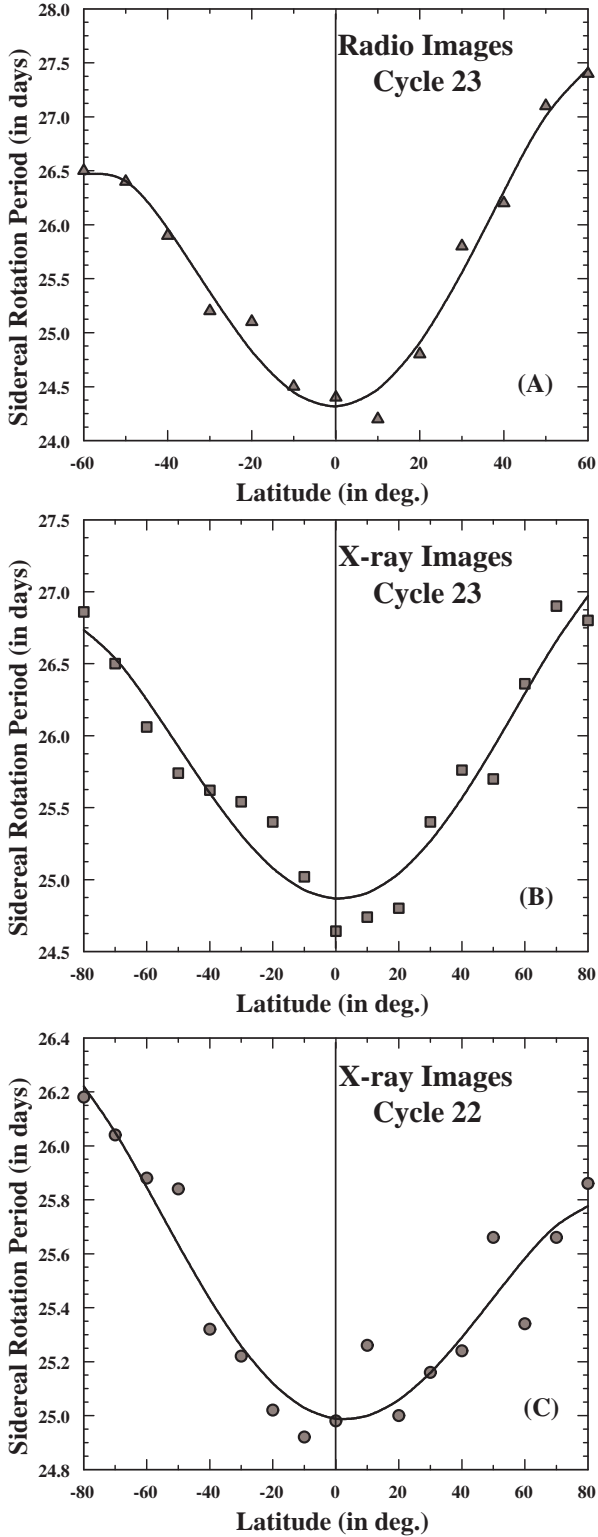}}
\caption{Average profiles of sidereal rotation period for the solar cycles 22 and 23 obtained by radio and X-ray images.}
\label{Figure4_NS}
\end{figure}

\section{Variation of sidereal rotation period in the solar corona}

The sidereal rotation periods during 1999 - 2005 were obtained using radio images at 17 GHz and the values are presented in the form of space-time plot in Figure \ref{Figure1_NS}. The details of the procedure for obtaining the sidereal rotation periods are given in \citet{chandra09} and \citet{chandra}. Here, the latitudes vary from $-$60 degree (in the southern hemisphere) to 60 degree (in the northern hemisphere). The red colour represents the sidereal rotation period of 29 days (highest period, which would be the slowest rotation rate) and violet represents the smallest rotation period of 23 days (fastest rotation rate). The red colour is only prevalent in the northern hemisphere which indicates that rotation periods in this part of the solar corona are higher than those in the southern hemisphere. This is a clear evidence of North-South asymmetry in the coronal rotation period. The asymmetry was highest in the year 2003. All the years 1999 - 2005 are in the solar cycle 23. Thus Figure \ref{Figure1_NS} indicates that in solar cycle 23, the solar corona in the northern hemisphere rotates slower than that in the southern hemisphere.

Using flux modulation methods on \textit{Yohkoh} X-ray images, \citet{chandra10} obtained sidereal rotation periods for the years 1992 - 2001. The solar cycle 23 began in 1996 and hence the measurements before 1996 are of solar cycle 22. Thus, the space-time plots of the sidereal rotation period, for the epoch during 1996 - 2001 (shown in Figure \ref{Figure2_NS}) and during 1992 - 1995 (shown in Figure \ref{Figure3_NS}), belongs to the solar cycle 23 and 22, respectively. Here, the range of rotation period is shorter (24 - 28 days) than that from the radio images (23 - 29 days). This shows that equatorial rotation periods obtained by the radio images are lower than those obtained by the X-ray images. In the higher latitudes, it is opposite of this, \textit{i.e.}, higher when obtained by the radio images than the X-ray images. 

The space-time plots in Figure \ref{Figure2_NS} show a reasonable resemblance to those in Figure \ref{Figure1_NS}, in that the rotation period is higher in the northern hemisphere than in the southern hemisphere. The Figures \ref{Figure1_NS} and \ref{Figure2_NS} have common epoch (1999 - 2001). From the careful visual inspection of the latitude range $\pm$60 degrees for both patterns, it can be imphasized that in this epoch  behavior of rotation rate and North-South asymmetry is reasonably similar. Both of these plots belong to the solar cycle 23. The rotation period in solar cycle 22 as seen in Figure \ref{Figure3_NS} also show clear evidence of N-S asymmetry, but in this solar cycle the rotation period is higher in the southern hemisphere than that in the northern hemisphere.

\begin{table*}
 \centering
 \begin{minipage}{134mm}
 \caption{Coefficients of Sidereal rotation rate $C_0$, $C_1$, $C_2$, $C_3$ and $C_4$ (in degree/day) obtained using radio images (NoRH at 17 GHz) and X-ray images (\textit{Yohkoh}/SXT). These values are used to derive the continuous curves of Figure \ref{Figure2_NS}.}
 \begin{tabular}{@{}lllcccccc@{}}

 \hline
  Data			&Solar	&	Data		& \multicolumn{5}{c}{Coefficients}				&Reference\\
  Source		&Cycle	&	Period	& $C_0$& $C_1$ & $C_2$ 	&	$C_3$ &	$C_4$ 	&			\\
 \hline
  
  NoRH/17GHz&	23  &1999-2005 	&14.821&$+0.030$&$-3.220$&$-0.417$&$+1.670$&Figure 2(A)\\
  Yohkoh/SXT&	23	&1996-2001 	&14.489&$+0.162$&$-0.991$&$-0.251$&$-0.101$&Figure 2(B)\\
  Yohkoh/SXT&	22	&1992-1995 	&14.406&$-0.022$&$-0.461$&$+0.154$&$-0.113$&Figure 2(C)\\
  
 \hline
\end{tabular}
\label{Table1_NS}
\end{minipage}
\end{table*}

The average profiles for the coronal rotation based on these measurements are shown in Figure \ref{Figure4_NS}. Figure \ref{Figure4_NS} has three panels (A, B and C): panel C for solar cycle 22 is the average of X-ray measurements during 1992 - 1995; the other two panels (A and B) are for solar cycle 23 from radio images at 17 GHz during 1999 - 2005 and from X-ray images during 1996 - 2001 respectively. The continuous curves (in Figure \ref{Figure2_NS}) represent 4th degree polynomials for the average sidereal rotation period, $T(\psi)$. The equation for such a polynomial is given as below:

\begin{equation}
\Omega(\psi)=C_0 + C_1 {\sin \psi} + C_2 {\sin^2\psi} + C_3 {\sin^3\psi} + C_4 {\sin^4\psi}
\label{Eq1_NS}
\end{equation}

where $\Omega(\psi)$ and $\psi$ are the sidereal rotation rate (in degree/day) and solar latitude (in degree), respectively. Whereas, $C_0$, $C_1$, $C_2$, $C_3$ and $C_4$ used in equation \ref{Eq1_NS} are the polynomial coefficients. The values of these coefficients for radio and X-ray measurements are given in Table \ref{Table1_NS}. The sidereal rotation period $T(\psi)$ can be obtained easily, knowing sidereal rotation rate $\Omega(\psi)$, using the following equation, 

\begin{equation}
T(\psi)=\frac{360}{\Omega(\psi)}
\label{Eq2_NS}
\end{equation} 

The values of $C_0$ indicate that the radio images estimate the coronal sidereal rotation rate somewhat higher than X-ray images. The N-S asymmetry will be through the coefficients $C_1$ and $C_3$. These coefficients have opposite signs for the solar cycles 22 and 23.

The average sidereal rotation period from the radio images during solar cycle 23 (Figure \ref{Figure4_NS}A) has a very clear asymmetry in North and South hemispheres. The rotation period at 60 degree North and South hemispheres differs by one day. This difference is about 4\%. The results of X-ray measurements for solar cycle 23 do support the radio measurements; however, the difference in North and South hemispheres is less. This could be partly due to the epoch difference and also X-ray measurements may represent higher heights in the corona than 17 GHz radio emission. The X-ray measurements of solar cycle 22 clearly have asymmetry in the opposite sense, wherein the southern hemisphere seems to have higher rotation period than the northern hemisphere, the highest difference being $>$ 0.5 day. This difference is about 2\%. The epochs in all the three panels, here, do not cover either solar cycle 22 or 23 completely. The results are indicative of the fact that the North-South asymmetry is real and significant. Moreover, the N-S asymmetry also changes its sign from even (22) to odd (23) solar cycles.

\section{Discussions}

The study of differential rotation and its temporal and spatial variation in the solar atmosphere is an important task of solar physics. Various solar observations have been used for such investigations. \citet{javaraiah05} showed that the sunspot cycles are connected by long-term variations in the equatorial rotation rate and the latitudinal gradient. \citet{song05} established, from the photospheric magnetic field measurements, that the degree of the asymmetry during the minimum of solar activity is higher than that during the maximum of solar activity. Moreover, the change of magnetic flux is always accompanied by a gradual shift of dominance from the northern hemisphere in the ascending phase to the southern hemisphere in the descending phase \citet{gigolashvili05} investigated H$_\alpha$ filaments and showed, statistically, the existence of N-S asymmetry in the chromospheric rotation during solar activity cycles 19 - 22. With the development of a cycle of solar activity, the N-S asymmetry of solar rotation rate changes, \textit{i.e.}, the differences in the velocity reach peak values at about solar activity maxima, but change sign from cycle to cycle near activity minima. Changes in the northern and southern hemispheres can be interpreted as oscillations with a 22-year period. The present study supports this finding of N-S asymmetry in the solar coronal rotation that (1) in odd solar cycles (\textit{e.g.} 23) the southern hemisphere rotates faster than the northern hemisphere and (2) in even solar cycles (\textit{e.g.} 22) the northern hemisphere rotates faster than the southern hemisphere. It can be concluded that the North-South asymmetry in solar corona and chromosphere is in phase. The rotational asymmetry probably leads to the manifestation of N-S asymmetry in many solar features observed by various methods and is certainly, in a complex manner, related to the solar dynamo.

\section*{Acknowledgments}

The authors wish to acknowledge the data (X-ray images from \textit{Yohkoh} mission and Nobeyama radio images at 17GHz) used in this work. These are acquired from the webpage of \textit{Yohkoh Data Archive Centre} (YDAC) and \textit{Nobeyama Radio Observatory} (NRO). We are indebted to the observers who were involved in the acquisition of the useful solar data. One of us (HOV) had detailed discussions on this work at 5 solar research groups in USA during August to October 2010. These discussions were very help in formulating this paper. The valuable suggestions by the referee and Dr Sachindra Naik are highly appreciated. The research at Physical Research Laboratory (PRL) is supported by Department of Space, Government of India. The authors also acknowledge the support given by their parent organizations during the course of this work.

\bsp

\label{lastpage}

\end{document}